\documentclass[twocolumn,pre,showpacs]{revtex4}
\usepackage{amssymb, amsmath, graphicx}
\def\be{\begin{equation}}
\def\ee{\end{equation}}

\begin{document}

\title{A Directed Network of Greek and Roman Mythology}

\author{Yeon-Mu \surname{Choi}$^{1}$}\email{ymchoi@mju.ac.kr}
\author{Hyun-Joo \surname{Kim}$^{2}$}\email{hjkim21@knue.ac.kr}
\affiliation{
$^1$ Center for Liberal Arts and Instructional Development,
 Myongji University, Kyongi-Do 449-728, Korea\\
$^2$ Department of Physics Education, Korea National University of Education,
 Chungbuk 363-791, Korea}

\date{\today}

\begin{abstract}
We study the Greek and Roman mythology using the network theory. 
We construct a directed network by using a dictionary of Greek and Roman 
mythology in which the nodes represent the entries listed in the dictionary
and we make directional links from an entry to other entries that appear 
in its explanatory part. We find that this network is clearly not a random 
network but a directed scale-free network. Also measuring the various
quantities which characterize the mythology network, we analyze the Greek 
and Roman mythology and compare it to other real networks.

\end{abstract}

\pacs{89.75.Fb, 89.75.Hc, 89.65.Ef}

\maketitle
  
 Current intense researches are focused on the widespread complex 
systems such as biological, social, technological, economic, 
and communications systems \cite{Albe, Newm, Buch, Watt, Doro, Past}. 
Such complex systems can be represented by the complex network which is 
composed of nodes which represent the diverse elements in the system and 
links by which the elements are connected if there are various interactions 
between them. For example, 
in a social acquaintance network nodes are individuals, links are connections 
between friends. In the World-Wide-Web(WWW) network nodes are web pages, links 
are hyper-links between them. Especially, in the WWW network, hyper-links 
have directions so it is called a directed network.

 The complex networks are characterized by some topological and
geometrical properties such as small-world, high degree of
clustering, and scale-free topology. The small-world property
denotes that the average path length $L$ which is the average
shortest path length between vertex pairs in a network, is small.
It grows logarithmically with the network size $N$. The
clustering structure in a network is measured by the clustering
coefficient $C$ which is defined as the fraction of pairs between the
neighbors of a vertex that are the neighbors  of each other. The
high degree of clustering  indicates that if vertices A and B are
linked to vertex C then A and B are also likely to be linked to each
other.  The scale-free (SF) topology
reflects that the degree distribution $P(k)$ follows a power law,
$P(k) \sim k^{-\gamma}$, where degree $k$ is the number of edges
attached to a vertex and $\gamma$ is the degree exponent.
Such network is called the SF network in which there are vertices of
high degree which produces strong effects.
Also recent attention has been focused on the hierarchical
structure and the cyclic topology.
A hierarchical
structure appears in some real networks and it has been clarified by
a power-law behavior of the clustering coefficient $C(k)$ as a
function of the degree $k$ \cite{ck1}. This
indicates that the networks are fundamentally modular. It is an   
origin of the high degree of clustering of the networks.
The cyclic topology is determined by loops with various sizes which 
can affect the delivery of information, transport process, and 
epidemic spreading behavior \cite{es}. The cyclic coefficient $R$
which considers the loops of all sizes  from three up to infinity 
is defined as the average of the local cyclic coefficient $r_i$ 
over all the vertices \cite{cc}. 
A local cyclic coefficient $r_i$ for a vertex $i$  is defined as the 
average of the inverse size of the smallest loop that connects the vertex $i$
and its two neighbor vertices, i.e., 
$ r_i = \frac{2}{k_i (k_i -1)} \sum_{<lm>} \frac{1}{S_{lm} ^i}$
where $<lm>$ is for all the pairs of the neighbors of the vertex $i$
and $S_{lm} ^i$ is the smallest size of the closed path that passes through 
the vertex $i$ and its two neighbor vertices $l$ and $m$.
The cyclic coefficient has a value between
zero and $1/3$. $R$=0 means that the network has a perfect tree-like
structure without  having any loops. Meanwhile if all the neighbor
pairs of the vertices have direct links  to each other, then  the
cyclic coefficient becomes $R$=1/3. The larger the cyclic coefficient $R$
is, the more cyclic the network is.

Applications of the network analysis to the real systems have 
gradually broaden from the systems whose network structures are obviously 
exposed, for examples, trains \cite{Stat}, subways \cite{Bost}, airports
\cite{Airp}, and internet communities \cite{Dati}, to the systems whose 
network structures are relatively hidden, for examples, language
\cite{Canc}, seismology \cite{Afte}, jazz \cite{Comm}, tango \cite{Tang}, 
comics \cite{Marv}, dolphins \cite{Dolp}. The complex network analysis 
provides a new successful standpoint to a wide range of real-life systems. 
Reversely, the studies of real-life networks are provided as basic data to 
construct the general theory of complex networks. Thus the studies for a 
variety of real-world systems using network analysis is very important.

In this paper, we study the Greek and Roman mythology (GRM) which is a 
reflection of human-life using the network analysis. 
By using a GRM dictionary we construct a directed mythology network in which 
the nodes correspond to myth characters and two nodes are linked when the 
corresponding characters have a relationship in the myth. 
We analyze the GRM by surveying the various properties for the  
GRM network and compare the myth-world network to other real-world networks.

A mythology comprises a collection of stories 
belonging to a single culture group which frequently feature both 
anthropomorphic or theriomorphic divine figures, heros, or animals.
The GRM becomes the patterns upon which Freudian 
psychiatrists base their interpretation of human behavior; painters, 
composers, sculptors and writers, deliberately or unconsciously, 
imitate the mythical patterns of the past.
Myth furnishes us more than a repertoire of literary plots and themes.
The GRM has been recounted by hundreds of writers throughout 
the world over the course of nearly three thousand years. 
It incorporates vast 
myth-tale motifs in which so many myth characters are connected each other.
The mythic world constitute a mythic-social network composed of 
myth characters and connections between them.

\begin{figure}
\includegraphics[width=7cm,height=4cm]{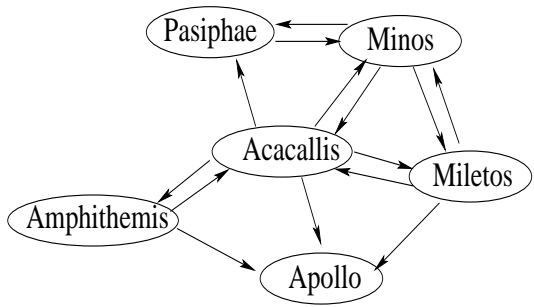}
\caption{The connection graph of Acacallis who has five neighbors. 
Its out-degree $k_{out}=5$ and
in-degree $k_{in}=3$.}\label{pic}
\end{figure}

The dictionary of classical mythology by M. Grant and J. Hazel \cite{Gran} 
is a biographical dictionary including the extensive Greek and Roman 
mythological 
characters, from major deities to the lesser-known nymphs. It includes, as 
$1647$ entries, gods, heros, monsters, mortals, fairies. Each entry has 
an biographical explanatory part in which other entries are referred to. In the 
mythology network we construct, the nodes correspond to the entries of the 
dictionary, and an outgoing link is formed from an entry to the other which 
appears in the explanatory part of the former, while an incoming link is 
formed from an entry to the entry referred by the former. 
That is, outgoing links of 
'refer to' and incoming links of 'referred by' on the dictionary are made. 
The entry Acacallis, as an example, has the explanatory part as "Daughter 
of Minos and Pasiphae. She bore Apollo a son, Amphithemis, and perhaps 
Miletos also." Acacallis has directed links to the five entries, Minos, 
Pasiphae, Apollo, Amphithemis, and Miletos.  Also, if the linked entries 
refer to 
Acacallis in their explanatory part opposite-directed links are formed between 
them.  
\begin{figure}
\includegraphics[width=7cm]{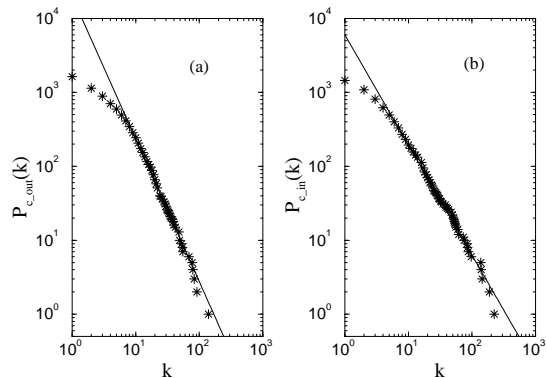}
\caption{
(a) The cumulative out-degree distribution of the GRM network.
 The slope of the solid line is $1.93$, which shows $\gamma_{out}=2.93$.
(b) The cumulative in-degree distribution of the GRM network. 
The slope of solid line is $1.49$,
which shows $\gamma_{in}=2.49$.} \label{bide}
\end{figure}
Figure $\ref{pic}$ shows the connection of Acacallis to its five neighbors.
There are bi-directional links between Acacallis and Minos, Miletos, and 
Amphithemis, while links between Acacallis and Pasiphae and Apollo are 
mono-directional. It results from that the explanatory parts of Minos, 
Miletos, and Amphithemis refer to Acacallis, while those of Pasiphae and 
Apollo do not refer to Acacallis. In this way, the node Acacallis has 
in-dgree $k_{in} = 3$ and out-degree $k_{out} = 5$ which are the number 
of links incoming upon it and outgoing from it, respectively.
Thus we construct the directed mythology network where the number of nodes, 
total in-links (out-links), and total undirected links are $1637$, $6687$ 
and $8938$ respectively. Here total undirected links are counted in the 
undirected network where we neglect the direction of links so two entries
are connected if they refer to or refer by each other.
The difference between the numbers of total undirected links and 
total directional links is same as the number of unilateral references 
in the dictionary. 
\begin{figure}
\includegraphics[width=7cm]{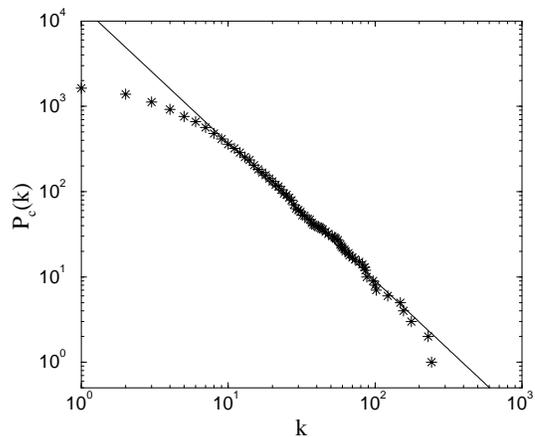}
\caption{ The cumulative degree distribution of the GRM network ignoring 
the directions of the edges. 
The slope of the solid line is $1.61$, which shows $\gamma=2.61$. } 
\label{dege}
\end{figure}

We measure the degree distribution of the GRM network
by using the cumulative degree distribution function. The cumulative degree 
distribution function $P_c (k)$ is the probability that a randomly selected 
node has over $k$ links, i.e., $ P_c (k) = \sum_{k'\geq k}^{\infty} P(k')$ .
The cumulative distribution also follows power-law, but with exponent 
$\gamma - 1$ rather than exponent $\gamma$, so that  
$P_{c} (k) \sim k^{-\gamma+1}$.
We measured the out-degree, the in-degree, and the total-degree distributions. 
The plots of the cumulative out-degree  and in-degree distribution functions 
as a function of the out-degree and the in-degree are shown in Fig.$\ref{bide}$
(a) and (b), respectively.   
The slopes of the straight guide lines are $1.93$ and $1.49$, which represents 
that the out-degree and the in-degree exponents are 
$\gamma_{out}$ $\approx$ $2.93$
and $\gamma_{in}$ $\approx$ $2.49$. Thus we found that the GRM network 
is a directed scale-free network, which means that all characters do not play 
an equal role and some characters play an more central role than other 
characters in the Greek and Roman mythic world. Also we obtained the values of 
degree exponent $\gamma$ $\approx$ $2.61$ in the undirected network 
as shown in Fig. $\ref{dege}$.

\begin{table}
\caption{Seven most connected entries with the corresponding out-degrees 
and in-degrees and degrees.}\label{tb}
\begin{ruledtabular}
\begin{tabular}{clrlrlr}
rank & & & entries & \\ 
 & &$k_{out}$ & &  $k_{in}$ & & $k$\\
\hline
1&Heracles& 140 &Zeus     &223   &Zeus     &243\\
2&Poseidon& 92  &Heracles &187   &Heracles &230\\
3&Odysseus& 83  &Apollo   &144   &Poseidon &177\\
4&Zeus    &79   &Nymph    &140   &Apollo   &156\\
5&Argonaut&77   &Poseidon &137   &Nymph    &149\\
6&Theseus &69   &Odysseus &97    &Odysseus &122\\
7&Dionysus&55   &Athena   &87    &Hera     &102\\

\end{tabular}
\end{ruledtabular}
\end{table}

Table $\ref{tb}$ shows seven most connected entries with the corresponding 
numbers of out-degrees, in-degrees, and undirected-degrees  in order of ranks.
We found that Heracles ranks the first with $140$ outgoing links for  
out-degrees, while Zeus ranks the first with $223$ incoming links and 
$243$ undirected links for in-degrees and undirected-degrees. 
In the GRM network, the fact that a nodes has more links than other nodes 
means that its corresponding myth character appears more frequently in myth 
tales. Heracles has the most outgoing links, which  means that he appears 
as a leading character in many different myth tales. On the one hand Zeus 
who has most incoming links most frequently appears as a supporting 
character in different myth tales. 
Also we notice that the characters well known to the mass of the people
hold high ranks on the whole. 

\begin{figure}
\includegraphics[width=7cm]{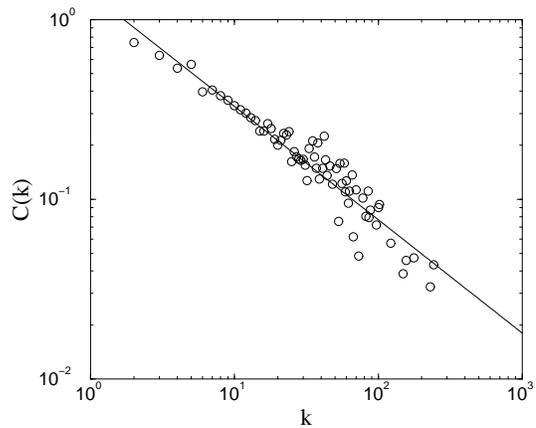}
\caption{The plot of the clustering coefficient $C(k)$ as a function of 
degree $k$ for the GRM network. 
The slope of the solid line is $0.63$. } \label{cluk}
\end{figure}

For the undirected GRM network, we measure various quantities which 
characterize the network.
First, we obtained the average path length $L=3.47$ and the 
clustering coefficient $C=0.41$. Comparing these values with the random 
network of the same number of nodes and links for which the values of the 
average shortest path length and  the clustering coefficient are $6.86$
and $0.0034$, we find that the GRM network has the 
small-world property and high degree of clustering as the other various complex 
networks. The small-world effect also underlies some well-known parlar games,
particularly the calculation of Erd\"os numbers \cite{Erno}. Similarly,
we measured Heracles numbers and Zeus numbers of the nodes. 
Heracles(Zeus) number of a node is defined as the shortest path length 
between the node and Heracles(Zeus). The Heracles(Zeus) numbers range from 
$0$(himself) to $7$($6$) and the average Heracles(Zeus) number of the network 
is $2.18$($2.11$). 

\begin{figure}
\includegraphics[width=7cm]{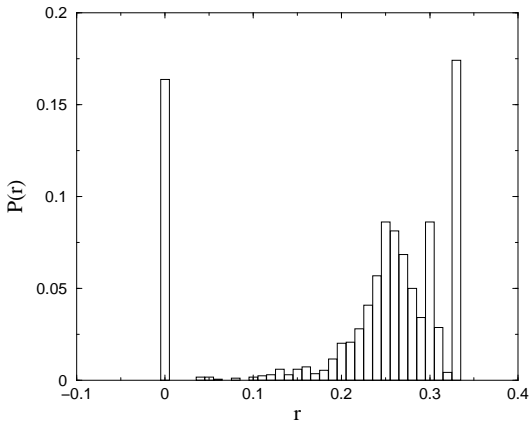}
\caption{
The distribution of local cyclic coefficient for the GRM network.
}
\label{sd}
\end{figure}

 Figure $\ref{cluk}$ shows the log-log plot of the clustering 
coefficient $C(k)$ versus the degree $k$. The straight guide line represents 
that $C(k)$ follows a power-law, $C(k) \sim k^{-\beta}$ with 
$\beta \approx 0.63$. 
It means that the GRM network forms a hierarchical structure. We also
measured the cyclic coefficient $R$ and obtained $R \approx 0.23$. Figure 
$\ref{sd}$ shows the plot of the distribution of local cyclic coefficient. 
There are the first and second peaks at $r=0$ and $r=1/3$ and two peaks have
almost equivalent values. This result represents that there are many tree-like 
and triangular patterns in the GRM networks. It is different from the
network structures of the other real-networks where the only one between 
tree-like 
and triangular pattern is certainly dominant \cite{cc}. While, except for 
$r=0$, it is similar to that of the movie actor network \cite{moac} in which 
nodes are actors
and two nodes are linked if the corresponding actors have acted in the same
movie together. 
It reflects the biographical nature of the mythology dictionary: two entries 
have high possibilities of appearance in the explanatory parts of each other 
when the corresponding characters have jointly appeared in the same myth 
story as two actors have costarred in a same film.
That is, although the GRM network is constructed by using a biographical 
dictionary, the GRM network can be regarded as a kind of a social network.

In summary, we studied the relationship among characters appeared in 
the GRM  with the help of the latest complex network theory. By using the 
biographical dictionary of GRM, we constructed the directed GRM network 
in which the nodes correspond to the entries(mythology characters) and
a directional link was made from an entry A to an entry B when the entry 
B was appeared in the explanatory part of the entry A. It was founded
that the GRM network is a scale-free network and has properties such as
the small-world and high degree of clustering. Also by measuring the 
clustering coefficient $C(k)$, we found that the GRM network forms a 
hierarchical structure. The distribution of local cyclic coefficient 
tell us that the GRM network is a social-like network such as the  
movie actor network as well as a dictionary network. 

This work was supported by the 2005 research grant from
Korea National University of Education.

\end{document}